\documentclass[12pt,14paper]{article}
\usepackage[english]{babel}
\usepackage[latin1]{inputenc}
\usepackage[normalem]{ulem}
\usepackage[intlimits]{amsmath}
\usepackage{amssymb}
\usepackage{geometry}
\usepackage{color}
\usepackage{verbatim} 
\usepackage{slashed}
\usepackage{rotating}
\usepackage{enumitem}

\begin{document}

\title{\vspace{-2cm} 
{\normalsize
\vspace{-.4cm}\flushright TUM-HEP 958/14\\
\vspace{-.4cm}\flushright FLAVOUR(267104)-ERC-84\\}
\vspace{0.6cm} 
\bf Lepton parameters in the see-saw model \\ extended  by one extra Higgs doublet
 \\[8mm]}

\author{Alejandro Ibarra$^1$  and Ana Solaguren-Beascoa$^{1,2}$ \\[2mm]
{\normalsize$^1$ \it Physik-Department T30d, Technische Universit\"at M\"unchen,}\\[-0.05cm]
{\normalsize \it James-Franck-Stra\ss{}e, 85748 Garching, Germany}\\[-0.05cm]
{\normalsize $^2$\it Max-Planck-Institut f\"ur Physik (Werner-Heisenberg-Institut),}\\ [-0.05cm]
{\normalsize \it F\"ohringer Ring 6, 80805 M\"unchen, Germany}
}
\maketitle

\begin{abstract}
We investigate the  radiative generation of lepton masses and mixing angles in the Standard Model extended by one right-handed neutrino and one extra Higgs doublet. We assume approximate rank-1 Yukawa couplings at a high energy scale and we calculate the one loop corrected charged lepton and neutrino mass matrices at the low energy scale. We find that quantum effects generate, for typical high energy parameters, a hierarchy between the muon and the tau mass, a hierarchy between the solar and  the atmospheric mass splittings, and a pattern of leptonic mixing angles in qualitative agreement with experiments. 
\end{abstract}

\section{Introduction}

The origin of the fermion mass hierarchies and mixing angles remains as one of the biggest mysteries in Particle Physics. Moreover, the progress in the determination of the neutrino mass splittings and mixing angles over the last decade, far from illuminating the mystery, has drawn a picture showing striking differences between the neutrino parameters and the quark or charged lepton parameters~\cite{Fogli:2012ua,GonzalezGarcia:2012sz}. The puzzle is three-fold: i) neutrino masses are much smaller than the quark and charged lepton masses, ii) the mass hierarchy in the neutrino sector is milder than in the quark and charged lepton sectors, iii) the entries of the leptonic mixing matrix are all ${\cal O}(0.1)$ while the entries of the quark mixing matrix display the strong hierarchy $|V_{ub}|, |V_{cb}|\ll |V_{us}|$.

A plausible explanation for the small neutrino masses compared to the quark or the charged lepton masses consists in introducing heavy Majorana right-handed neutrinos, much heavier than the electroweak symmetry breaking scale~\cite{seesaw}. This is the renown see-saw mechanism. In this framework, the overall neutrino mass scale parametrically depends on the square of the neutrino Yukawa coupling and on the inverse of the right-handed neutrino mass. Therefore, the see-saw mechanism makes the generic prediction that neutrino masses should be much smaller than the quark or charged lepton masses, although the precise value cannot be predicted. Furthermore, and due to the high scale of lepton flavour violation, this framework predicts tiny rates for the rare lepton decays~\cite{meg_SM}, in agreement with the stringent experimental bounds ${\rm BR}(\mu\rightarrow e\gamma)\leq 5.7 \times 10^{-13}$~\cite{Adam:2013mnn}, ${\rm BR}(\tau\rightarrow \mu\gamma)\leq 4.4\times 10^{-8}$~\cite{Aubert:2009ag}, ${\rm BR}(\tau\rightarrow e\gamma)\leq 3.3\times 10^{-8}$~\cite{Aubert:2009ag}.

Whereas the seesaw model is a compelling framework to explain the smallness of neutrino masses, it tends to generate too large neutrino mass hierarchies~\cite{Casas:2006hf}. This drawback of the see-saw mechanism is automatically cured by introducing a second Higgs doublet. As discussed in \cite{Ibarra:2011gn,Grimus:1999wm}, even if the tree level neutrino masses display very large hierarchies,  quantum effects  induced by the second Higgs doublet generate a mild hierarchy between the heaviest and next-to-heaviest neutrino   masses which is, for   well motivated choices of the high energy parameters, in qualitative agreement with the value inferred from oscillation experiments. 

The generation of the mass hierarchies for the quark sector in the two Higgs doublet model was investigated in \cite{Ibarra:2014fla}. Assuming  tree level Yukawa couplings with very hierarchical eigenvalues (such that they can be considered in practice as rank-1), quantum effects generate, for generic high energy parameters, a mass hierarchy between the second and third generations which is of ${\cal O}(0.01)$, in qualitative agreement with the data. Moreover, assuming aligned tree level quark mass matrices, quantum effects generate  $|V_{us}|={\cal O}(0.1)$ and $|V_{ub}|, |V_{cb}|\ll |V_{us}|$, also in qualitative agreement with the measured values.

It was also pointed out in \cite{Ibarra:2014fla} that a similar mechanism would generate radiatively a muon mass in the presence of right-handed neutrinos (see also \cite{Joaquim:2014gba}). In this paper we carefully explore the generation of charged lepton and neutrino masses, as well as the leptonic mixing angles, in a simple scenario where the Standard Model is extended by right-handed neutrinos and one extra Higgs doublet. In section \ref{section:tree-level} we present the model and we calculate the lepton masses and mixings in a simplified scenario with just one right-handed neutrino and where all Yukawa couplings are rank-1 at the cut-off scale, hence only one generation has  non-vanishing tree-level masses. Then, in section \ref{section:quantum-effects} we discuss the impact of quantum effects on the leptonic parameters, finding rank-2 charged lepton and neutrino mass matrices. In section \ref{sec:LFV} we briefly address the generation of leptonic rare decays in this framework and, finally, in  section \ref{section:conclusions} we present our conclusions.

\section{The two Higgs doublet model extended with right-handed neutrinos: Tree level results}
\label{section:tree-level}

We consider in this paper an extension of the Standard Model consisting in adding one extra Higgs doublet and at least one right-handed neutrino, singlet under the Standard Model gauge group. We do not impose any  global or discrete symmetry on the model. Then, the flavour dependent part of the leptonic Lagrangian is given by:
\begin{equation}
-{\cal L}^{lep}=
Y_{e,ij}^{(a)} \bar l_{Li} e_{Rj} \Phi_a 
+Y_{\nu,ij}^{(a)} \bar l_{Li} \nu_{Rj} \tilde \Phi_a 
-\frac{1}{2} M_{{\rm M},ij} \bar\nu^C_{Ri}\nu_{Rj}+{\rm h.c.}
\end{equation}
where $i,j=1,2,3$  are flavour indices, $a=1,2$ is a Higgs index and $\tilde{\Phi}_a=i\tau_2\Phi^*_a$. It is convenient to work in the Higgs basis where one of the Higgs fields, say $\Phi_2$, does not acquire a vacuum expectation value. Therefore $\langle \Phi^0_1 \rangle =v/\sqrt{2}$, with $v=246$ GeV,  and $\langle \Phi_2^0\rangle=0$. In this basis, the charged lepton and  Dirac neutrino masses are proportional to the Yukawa couplings $Y_e^{(1)}$ and $Y_\nu^{(1)}$, respectively. Furthermore, we choose to work in the basis where the charged lepton Yukawa coupling to the Higgs $\Phi_1$, $Y_e^{(1)}$, is diagonal and real.  

We assume that the mass scale of the right-handed neutrinos is much larger than the electroweak symmetry breaking scale and the mass of all the extra Higgs  states $H^0$, $A^0$, $H^\pm$, which we denote collectively by $m_H$. Hence, at the energies relevant to current experiments, the right-handed neutrinos are decoupled and the theory can be conveniently described by the following effective Lagrangian:
\begin{align}
-{\cal L}^{\nu,\rm eff} &=
Y_{e,ij}^{(a)} \bar l_{Li} e_{Rj} \Phi_a +
\frac{1}{2}\kappa_{ij}^{(ab)} 
(\bar l_{Li} \tilde \Phi_a) 
(\tilde \Phi_b^T  l^C_{Lj}) +{\rm h.c.}
\end{align}
where, at the scale of the lightest right-handed neutrino mass, the coefficients of the dimension 5 operators $\kappa^{(ab)}$ read:
\begin{align}
\kappa^{(ab)}(M_1)=(Y_\nu^{(a)} M_{\rm M}^{-1} Y_\nu^{(b)T})(M_1)\;.
\label{eq:kappaab}
\end{align}
Hence, with our choice of basis for the leptonic and the Higgs fields, the neutrino mass matrix at the scale of the lightest right-handed neutrino mass depends just on the coupling $\kappa^{(11)}$:
\begin{align}
{\cal M}_\nu(M_1)=\frac{v^2}{2}\kappa^{(11)}(M_1)\;,
\end{align}
which is diagonalized in the standard way:
\begin{align}
{\cal M}_\nu=U^* {\rm diag}(m_1,m_2,m_3)U^\dagger\;.
\end{align}

We consider in what follows a scenario with one right-handed neutrino with mass $M_{\rm maj}$ and where all the Yukawa couplings are rank-1 at the cut-off scale $\Lambda$, with $\Lambda> M_{\rm maj}$. By an appropriate choice of the basis for the leptonic fields, it can be checked that the Yukawa couplings to the Higgs $\Phi_1$ can be written as:
\begin{align}
Y_e^{(1)}(\Lambda)=
\begin{pmatrix} 
0 & 0& 0 \\ 0 & 0 & 0 \\  0 & 0 & y_e^{(1)}
\end{pmatrix}\;,
\hspace{.2cm}
Y_\nu^{(1)}(\Lambda)=y_\nu^{(1)}
\begin{pmatrix} 
0 \\ \sin \alpha \\  \cos \alpha 
\end{pmatrix},
\end{align}

On the other hand, the charged lepton Yukawa coupling to the Higgs $\Phi_2$ must take the most general form of a rank-1 matrix, namely:
\begin{align}
Y_e^{(2)}(\Lambda)=
E_L^\dagger
\begin{pmatrix} 
0 & 0& 0 \\ 0 & 0 & 0 \\  0 & 0 & y_e^{(2)}
\end{pmatrix}
E_R\;, 
\label{eq:Ye2}
\end{align}
where $E_{L,R}$ are $3\times3$ unitary matrices. The Yukawa matrix elements are $Y^{(2)}_{e,ij}=y_e^{(2)}(E_L)^*_{3i} (E_R)_{3j}$,  hence only the last row of the unitary matrices is relevant, which we parametrize as:
\begin{eqnarray}
(E_L)_{31}&=&e^{i\rho_{e_L}}\sin\theta_{e_L}\sin\omega_{e_L}\;,\nonumber \\
(E_L)_{32}&=&e^{i\xi_{e_L}} \sin\theta_{e_L}\cos\omega_{e_L}\;,\nonumber\\
(E_L)_{33}&=&\cos\theta_{e_L}\;,
\end{eqnarray}
and similarly for $E_R$. Besides, the neutrino Yukawa coupling to the Higgs $\Phi_2$ must take the most general form of a column matrix:
\begin{align}
Y_\nu^{(2)}(\Lambda)=
y_\nu^{(2)} 
\begin{pmatrix} 
e^{i\rho_{\nu}}\sin\theta_{\nu}\sin\omega_{\nu} \\
e^{i\xi_{\nu}} \sin\theta_{\nu}\cos\omega_{\nu}\\
\cos\theta_{\nu}
\end{pmatrix}\;.
\end{align}
Analogous parametrizations for the quark Yukawa couplings can be found in \cite{Ibarra:2014fla}. In what follows we will neglect  all the phases for simplicity.

With this parametrization for the parameters at the cut-off scale $\Lambda$ it is straightforward to calculate the tree-level charged lepton and neutrino masses. The result is:
\begin{eqnarray}
&m_\tau|_{\rm tree}=\displaystyle{\frac{v}{\sqrt{2}} \, y_e^{(1)}},~~~~~~~ &m_\mu|_{\rm tree}=m_e|_{\rm tree}=0\\
&m_3|_{\rm tree}=\displaystyle{\frac{v^2}{2}\,\frac{y_\nu^{(1)2}}{M_{\rm maj}}}, ~~~~~~~& m_2|_{\rm tree}=m_1|_{\rm tree}=0
\end{eqnarray}
The leptonic mixing matrix, on the other hand, is not univocally defined since an internal rotation between the first two generations of charged lepton or neutrino fields leaves the Lagrangian invariant. As a result, only the 33 element of the mixing matrix is defined, which reads $U_{33}|_{\rm tree}=\cos\alpha$.

\section{Quantum effects on charged lepton and neutrino parameters}
\label{section:quantum-effects}

We consider now the  impact of quantum effects on the scenario presented in the previous section, where the Standard Model is extended with one extra Higgs doublet and one right-handed neutrino, under the assumption that all Yukawa couplings are rank-1 at a cut-off scale $\Lambda \gg M_{\rm maj}$. The one loop corrected Yukawa couplings at the scale of decoupling of the right-handed neutrinos approximately read:
\begin{align}
&Y_{e,ij}^{(a)}(M_{\rm maj}) \simeq Y_{e,ij}^{(a)}(\Lambda)+ \frac{1}{16\pi^2}\beta_{e,ij}^{(a)}(\Lambda)\log\frac{\Lambda}{M_{\rm maj}}\;,\nonumber\\ 
&Y_{\nu,i}^{(a)}(M_{\rm maj})\simeq Y_{\nu,i}^{(a)}(\Lambda)+ \frac{1}{16\pi^2}\beta_{\nu,i}^{(a)}(\Lambda)\log\frac{\Lambda}{M_{\rm maj}}\;,
\label{eq:one-loop-corrected}
\end{align}
where the beta-functions $\beta_{e}^{(a)}$, $\beta_{\nu}^{(a)}$ can be found in Appendix \ref{app:Beta-functions}. As discussed in \cite{Ibarra:2014fla}, the one-loop corrected Yukawa coupling $Y_{e}^{(1)}(M_{\rm maj})$ is rank-2, hence radiatively generating a non-vanishing muon mass.  Besides, as a consequence of the breaking of the mass degeneracy between the first- and second-generation  of leptons, the matrices that diagonalize the Yukawa coupling become unambiguously fixed. More specifically, casting  $Y_e^{(1)}=U_{e_L} {\rm diag}(y_e,y_\mu,y_\tau)U_{e_R}^\dagger$, we find for the  Yukawa eigenvalues,
\begin{eqnarray}
y_{\tau}(M_{\rm maj})&\simeq&y_e^{(1)}\;, \nonumber\\
y_\mu(M_{\rm maj})&\simeq&\frac{1}{16\pi^2}\left[\sum_{i,j=1}^2 \left(\beta_{e,ij}^{(1)}(\Lambda)\right)^2\right]^{1/2} \log\frac{\Lambda}{M_{\rm maj}} \;, \nonumber\\
y_e(M_{\rm maj})&=&0\;,
\end{eqnarray}
while for the unitary matrices $U_{e_L}$ and $U_{e_R}$,
\begin{equation}
U_{e_L}(M_{\rm maj})\simeq
\left(
\begin{array}{ccc}
 \cos\zeta_{e_L} &  \sin \zeta_{e_L} &  0 \\
 -\sin\zeta_{e_L} &  \cos\zeta_{e_L} & 0 \\ 
0 &  0 & 1 
\end{array}
\right),~~~~~~
U_{e_R}(M_{\rm maj})\simeq
\left(
\begin{array}{ccc}
 \cos\zeta_{e_R} &  \sin \zeta_{e_R} &  0 \\
 -\sin\zeta_{e_R} &  \cos\zeta_{e_R} & 0 \\ 
0 &  0 & 1 
\end{array}
\right)\;,
\label{eq:UeL-UeR}
\end{equation}
where
\begin{equation}
\tan \zeta_{e_L} =\left. \left(\frac{\beta^{(1)}_{e,11}\beta^{(1)}_{e,21}+\beta^{(1)}_{e,12}\beta^{(1)}_{e,22}}{(\beta^{(1)}_{e,21})^2+(\beta^{(1)}_{e,22})^2}\right)\right|_\Lambda, ~~
\tan \zeta_{e_R} =\left.\frac{\beta^{(1)}_{e,11}\beta^{(1)}_{e,12}+\beta^{(1)}_{e,21}\beta^{(1)}_{e,22}}{(\beta^{(1)}_{e,12})^2+(\beta^{(1)}_{e,22})^2}\right|_\Lambda\;.
\label{eq:zetas}
\end{equation}

At the scale $M_{\rm maj}$ we redefine the leptonic fields to make the Yukawa coupling  to the Higgs $\Phi_1$ diagonal, which we denote by $\tilde Y_e^{(1)}(M_{\rm maj})$. In this basis,
\begin{eqnarray}
\tilde Y_e^{(a)}(M_{\rm maj})&=& (U_{e_L}^\dagger Y_e^{(a)} U_{e_R})(M_{\rm maj})\;, 
\nonumber \\
\tilde Y_\nu^{(a)}(M_{\rm maj})&=& (U_{e_L}^\dagger Y_\nu^{(a)})(M_{\rm maj})\;.
\label{eq:Ys-newbasis}
\end{eqnarray}

The decoupling of the heavy right-handed neutrino generates dimension five operators, which can be calculated from matching the full theory to the effective theory at the scale $M_{\rm maj}$. The result is:
\begin{equation}
\kappa_{ij}^{(ab)}(M_{\rm maj})=\frac{1}{M_{\rm maj}} \left.\tilde Y^{(a)}_{\nu,i}\tilde Y^{(b)}_{\nu,j}\right|_{\rm maj}\;,
\end{equation}
which is rank 1 and therefore only has one non-vanishing eigenvalue. 

The Yukawa couplings ${\tilde Y}_e^{(a)}$ and the couplings $\kappa^{(ab)}$ of the dimension-5 operators are subject to quantum corrections between the scale $M_{\rm maj}$ and $m_H$; the corresponding renormalization group equations are also given in Appendix \ref{app:Beta-functions}. The charged lepton Yukawa couplings at the scale $m_H$  approximately read:
\begin{eqnarray}
y_\tau(m_H)&\simeq& y_e^{(1)}\nonumber \\
y_\mu(m_H)&\simeq&\frac{1}{16\pi^2}\left\{\left[\sum_{i,j=1}^2 \left(\beta_{e,ij}^{(1)}(\Lambda)\right)^2\right]^{1/2}
 \log\frac{\Lambda}{M_{\rm maj}} +\left[\sum_{i,j=1}^2 \left(\beta_{e,ij}^{(1)}(M_{\rm maj})\right)^2\right]^{1/2}
\log\frac{M_{\rm maj}}{m_H}\right\}
\;, \nonumber\\
y_e(m_H)&=&0\;,
\end{eqnarray}
while the unitary matrices $U_{eL}$, $U_{eR}$ at the scale $m_H$ are approximately equal to the corresponding matrices at the scale $M_{\rm maj}$, given by Eqs.~(\ref{eq:UeL-UeR},\ref{eq:zetas}).

On the other hand, the coupling $\kappa^{(11)}(M_{\rm maj})$ has two degenerate (vanishing) eigenvalues and  quantum effects can significantly alter the structure of the mass matrix. Indeed, as discussed in \cite{Ibarra:2011gn,Grimus:1999wm}, quantum  effects in the two Higgs doublet model increase the rank of this coupling, thus breaking the degeneracy among the eigenvalues and fixing univocally the angles of the mixing matrix. Using the results of \cite{Ibarra:2011gn}, one finds, again neglecting   subleading effects,
\begin{eqnarray}
m_3(m_H)&\simeq& \frac{ v^2}{2 M_{\rm maj}}\displaystyle{| \tilde Y_\nu^{(1)}|^2}\;, \\
m_2(m_H)&\simeq& \frac{1}{16\pi^2}\frac{\lambda_5  v^2}{M_{\rm maj}}
\left[ \displaystyle{| \tilde Y_\nu^{(2)}|^2}- \frac{\displaystyle{| \tilde Y^{(2)\dagger}_\nu \tilde Y^{(1)}_\nu|^2}}{\displaystyle{|\tilde Y^{(1)}_\nu|^2}}\right]
\log\frac{M_{\rm maj}}{m_H}\;, \\
m_3(m_H)&\simeq& 0\;,
\end{eqnarray}
where $\lambda_5$ is the coupling  constant in  the potential term $V\supset \frac{1}{2} \lambda_5 (\Phi_1^\dagger \Phi_2)^2$. Besides, the columns of the leptonic mixing matrix read:
\begin{eqnarray}
V_{\nu,i3}(m_H)  &\simeq& \frac{\tilde Y^{(1)}_{\nu i}}{\displaystyle{|\tilde Y^{(1)}_\nu|}}\;,\nonumber\\
V_{\nu,i2}(m_H)  &\simeq& \frac{1}{N_2}\left[\tilde Y^{(2)}_\nu -\frac{\tilde Y^{(2)\dagger} \tilde Y^{(1)}_\nu }{\displaystyle{|\tilde Y^{(1)}_\nu |^2}}\tilde Y^{(1)}_\nu \right]\;,
\end{eqnarray}
with $|\tilde Y^{(a)}_\nu |^2=\sum_i | \tilde Y^{(a)}_{\nu i}|^2$ and
\begin{equation}
N_2 \equiv \left[\tilde Y^{(2)\dagger}_\nu \tilde Y^{(2)}_\nu -\frac{\displaystyle{|\tilde Y^{(2)\dagger} \tilde Y^{(1)}_\nu |^2}}{\displaystyle{|\tilde Y^{(1)}_\nu |^2}}\right]^{1/2}\;,
\end{equation}
where the neutrino Yukawa couplings $\tilde Y^{(a)}$ are all evaluated at the scale $M_{\rm maj}$.

It is now straightforward to calculate the leading contributions to the charged lepton and neutrino masses, as well as to the leptonic mixing matrix, in terms of the parameters defining the Yukawa couplings at the cut-off scale, $Y_x^{(a)}(\Lambda)$,   $x=e,\nu,u,d$. The ratio between the muon and tau masses is approximately given by:
\begin{equation}
\frac{m_\mu}{m_\tau}\simeq \frac{1}{16 \pi ^2}\frac{y^{(2)}_e}{y^{(1)}_e}(P^2+Q^2)^{1/2}\;,
\label{eq:ratio-approx}
\end{equation}
where we have defined
\begin{eqnarray}
P&\equiv&
y_\nu ^{(1)}y_\nu ^{(2)} p_\nu  \log \left(\frac{\Lambda}{M_{\rm maj}}\right) +
\sum_{x=e,u,d}y_x ^{(1)}y_x ^{(2)} p_x \log \left(\frac{\Lambda}{m_H}\right) \;,\\
Q&\equiv&
y_\nu ^{(1)}y_\nu ^{(2)} q_\nu  \log \left(\frac{\Lambda}{M_{\rm maj}}\right) +
\sum_{x=e,u,d}y_x ^{(1)}y_x ^{(2)} q_x \log \left(\frac{\Lambda}{m_H}\right) \;,
\label{eq:def_p&q}
\end{eqnarray}
$p_x$, $q_x$  being complicated functions of the angles that diagonalize the Yukawa matrices $Y_x^{(2)}(\Lambda)$, and which are explicitly given in Appendix \ref{app:P&Q}. On the other hand, the ratio between the masses of the heaviest and the next-to-heaviest neutrino is:
\begin{equation}
\frac{m_2}{m_3}\simeq  \frac{\lambda_5}{8\pi^2}
\displaystyle{\left(\frac{y_\nu^{(2)}}{y_\nu^{(1)}}\right)^2} (1-\cos^2\psi)
\log\frac{M_{\rm maj}}{m_H}\;.
\end{equation}
Here, $\cos\psi$ measures the misalignment between the vectors $Y^{(2)}_\nu$ and $Y^{(1)}_\nu$ and is given in terms of the high energy parameters by $\cos\psi \equiv \cos\alpha\cos\theta_\nu+\sin\alpha \sin\theta_\nu \cos\omega_\nu$. Finally, we find for the mixing angles:
\begin{eqnarray}
\tan\theta_{12}&\simeq&\frac{\sin \zeta_{e_L} \cos \alpha  \left(\cos \alpha  \sin \theta _\nu  \cos \omega _\nu -\sin \alpha  \cos \theta _\nu \right)-\cos \zeta_{e_L} \sin \theta _\nu  \sin \omega _\nu }{\cos \alpha  \sin \theta _\nu  \left(\cos \zeta_{e_L} \cos \omega _\nu +\sin \zeta_{e_L} \sin \omega _\nu \right)-\cos \zeta_{e_L} \sin \alpha  \cos \theta _\nu }\;,\nonumber \\
\sin\theta_{13}&\simeq & - \sin \zeta_{e_L} \sin \alpha\;, \nonumber \\
\tan\theta_{23}&\simeq & \cos \zeta_{e_L} \tan \alpha \;,
\label{eq:mixing-angles}
\end{eqnarray}
where 
\begin{equation}
\tan \zeta_{e_L} = \frac{P}{Q}\;,
\label{eq:zetaeL}
\end{equation}
with $P$, $Q$ defined in Eq.~(\ref{eq:def_p&q}). In the case $\alpha=0$ one recovers the zero-th order structure of the CKM matrix derived in \cite{Ibarra:2014fla}. Therefore, in this framework the different pattern of mixing angles in the quark and lepton sectors is related to the amount of misalignment of the tree level (rank-1) Yukawa couplings to the Higgs $\Phi_1$, parametrized by one single angle $\alpha$. 

The ratio between the muon and the tau masses depends on various high-energy parameters, nonetheless, it is possible to estimate the typical size of this ratio for reasonable assumptions about the model parameters. For generic values of the angles that diagonalize the Yukawa couplings $Y_x^{(a)}$, one finds $p_x,q_x={\cal O}(0.1)$ for all $x=e,\nu,u,d$. Moreover, assuming $y^{(2)}_x\sim y^{(1)}_x$, we expect $P$ and $Q$ to be dominated by the up-type Yukawa couplings or, possibly, by the neutrino Yukawa coupling if the right-handed neutrino mass is sufficiently large. Note also that in the former case  the mass ratio is enhanced by a larger logarithm. Therefore,  and setting for concreteness $\log(\Lambda/M_{\rm maj})\sim10$, $\log(\Lambda/m_H)\sim 30$, we find a ratio between the muon and the tau mass approximately given by:
\begin{equation}
\frac{m_\mu}{m_\tau}\sim (0.01-0.1) \frac{y^{(2)}_e}{y^{(1)}_e}
{\max}\left\{y_\nu ^{(1)}y_\nu ^{(2)}, 3 y_u ^{(1)}y_u ^{(2)}\right\}\;.
\label{eq:mmumtau-approx}
\end{equation}

\begin{figure}
\begin{center}
\includegraphics[width=.49 \linewidth]{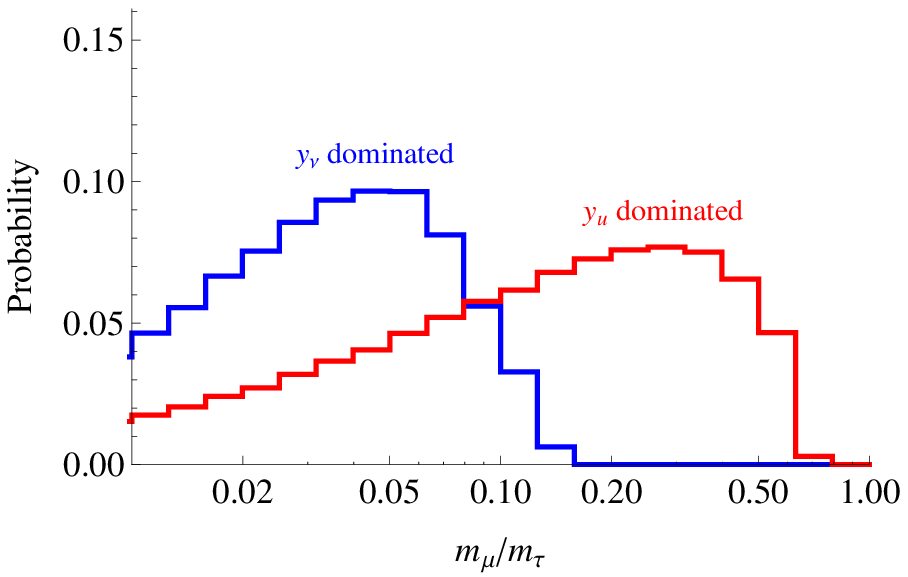}
\includegraphics[width=.49 \linewidth]{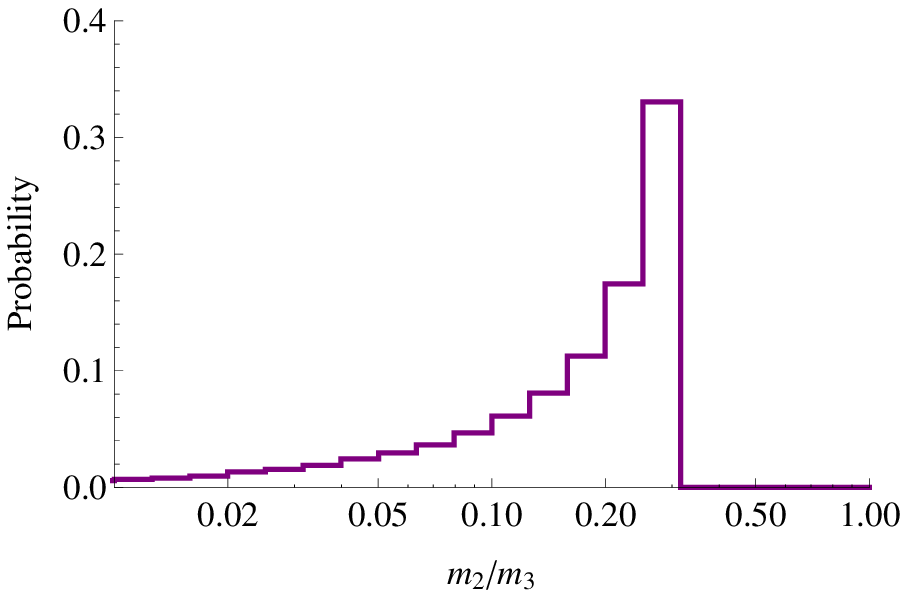}
\caption{Probability distributions of $m_\mu/m_\tau$ (left plot) and $m_2/m_3$ (right plot) from a random scan of the angles that diagonalize the rank-1 Yukawa matrices at the cut-off scale. See details in the text.}
\label{fig:distributions}
\end{center}
\end{figure}

We show in Fig.~\ref{fig:distributions}, left plot, the probability distributions of $m_\mu/m_\tau$ in a logarithmic binning from performing a random scan of the angles (with flat distributions between 0 and $2\pi$) and fixing for concreteness $y^{(2)}_e=y^{(1)}_e$, $\Lambda=10^{19}$ GeV, $M_{\rm maj}=10^{14}$ GeV and $m_{H}=10^{4}$ GeV. It was also assumed that the muon generation is dominated either by quantum effects induced by the right-handed neutrino (blue line) or by the top quark (red line); in each case we fix, respectively, $y_\nu ^{(1)}=y_\nu ^{(2)}=1$ or $y_u ^{(1)}=y_u ^{(2)}=1$. The probability distributions for other values of the Yukawa couplings can be straightforwardly determined with an appropriate rescaling of the horizontal axis of the figure. The large hierarchy measured between the muon and tau masses suggests that the radiative generation of the muon mass is dominated by the right-handed neutrino in the loop, due to the shorter renormalization group running. Nevertheless, the muon mass could also be generated by the top loop for appropriate choices of the Yukawa couplings $y^{(a)}_e$ and $y^{(a)}_u$ in Eq. (\ref{eq:mmumtau-approx}).

The order of magnitude of the neutrino mass ratio can be estimated along similar lines. For generic values of the high energy mixing angles and for typical mass scales of the model,  one finds $\log(\Lambda/M_{\rm maj})\sim 10$ and $\cos\psi={\cal O}(0.1)$. Hence,
\begin{equation}
\frac{m_2}{m_3}\simeq  (0.1-1) \,\lambda_5\,
\displaystyle{\left(\frac{y_\nu^{(2)}}{y_\nu^{(1)}}\right)^2}\;.
\end{equation}
Assuming again $y^{(2)}_\nu\sim y^{(1)}_\nu$ and  $\lambda_5={\cal O}(1)$, we find $m_2/m_3={\cal O}(0.1)$, also in qualitative agreement with experimental data. This is again confirmed by our numerical analysis, shown in Fig.~\ref{fig:distributions}, right plot, where we  present the probability distribution of $m_2/m_3$ in a logarithmic binning fixing for concreteness $y^{(2)}_\nu=y^{(1)}_\nu$, $\lambda_5=1$, $M_{\rm maj}=10^{14}$ GeV and $m_H=10^4$ GeV.

Lastly, concerning the expected values of the leptonic mixing angles in this framework, we first note that the angle $\zeta_{e_L}$ is generically neither maximal nor   small, as follows from Eq.~(\ref{eq:zetaeL}), thus leading to a sizable charged lepton contribution to the neutrino mixing. Then, assuming $\sin\alpha={\cal O}(0.1)$,  all three mixing angles are expected to be neither maximal nor small, also in qualitative agreement with global fits to neutrino oscillation experiments, which give as central values $\sin\theta_{12}\simeq 0.55$, $\sin\theta_{13}=0.15$ as $\sin\theta_{23}=0.64$  (assuming normal mass hierarchy) \cite{GonzalezGarcia:2012sz}.

The model contains a large number of high energy parameters, nevertheless it is possible to determine univocally the  parameters $\alpha$ and $\zeta_{e_L}$ in terms of the low energy observables $\sin\theta_{13}$ and $\tan\theta_{23}$. From Eq.~(\ref{eq:mixing-angles}) it follows that:
\begin{eqnarray}
\sin^2\alpha&\simeq& \frac{\sin^2\theta_{13}+\tan^2\theta_{23}}{1+\tan^2\theta_{23}}\;,\\
\sin^2\zeta_{e_L}&\simeq &\frac{\sin^2\theta_{13}(1+\tan^2\theta_{23})}{\sin^2\theta_{13}+\tan^2\theta_{23}}\;.
\end{eqnarray}
Adopting the central values of the mixing parameters one readily obtains $|\sin\alpha|\simeq 0.65$, $|\sin\zeta_{e_L}|\simeq 0.23$. Furthermore, the values of the angles $\theta_\nu$ and $\omega_\nu$ are correlated from the requirement of correctly reproducing the solar mixing angle. More specifically, we find:
\begin{equation}
\tan\theta_\nu=\frac{\sin\alpha (\tan\theta_{12}\cos\zeta_{e_L}-\cos\alpha \sin\zeta_{e_L})}{\tan\theta_{12}\cos\alpha\cos(\zeta_{e_L}-\omega_\nu)-\cos^2\alpha\cos\omega_\nu\sin\zeta_{e_L}+\cos\zeta_{e_L}\sin\omega_\nu}\;.
\end{equation}
The measured leptonic mixing angles can then be easily accommodated in this framework for generic high energy Yukawa structures, namely Yukawa matrices with mixing angles which are neither maximal nor  small.

\section{Lepton Flavour Violation}
\label{sec:LFV}

Any model of neutrino masses predicts a non-vanishing rate for the charged lepton flavour violating decays. In the Standard Model extended by right-handed neutrinos and one extra Higgs doublet, and working as usual in the basis where the charged lepton and the right-handed neutrino mass matrices are flavour diagonal, the Yukawa couplings $\tilde Y_e^{(2)}$ and $\tilde Y_\nu^{(a)}$ break lepton flavour and induce the rare decays $\ell_i\rightarrow \ell_j \gamma$ via quantum effects, suppressed by powers of the heavy Higgs mass or the heavy neutrino mass, respectively. In the framework of fermion mass generation discussed in this paper $M_{\rm maj}\gg m_H$, therefore we expect the leptonic rare decays to be dominated by diagrams involving the coupling $\tilde Y_{e}^{(2)}$. The value of this coupling at low energies approximately reads ${\tilde Y}_e^{(2)}(m_H) \simeq U_{e_L}^\dagger Y_e^{(2)} U_{e_R}$, where $Y_e^{(2)}$ was given in Eq.(\ref{eq:Ye2}) and $U_{e_L}, U_{e_R}$ in Eq.~(\ref{eq:UeL-UeR}). An explicit calculation yields $\tan\zeta_{eR}=\tan\omega_{eR}$, therefore the coupling $\tilde Y_{e}^{(2)}$ reads:
\begin{equation}
\tilde Y_e^{(2)}(m_H)
\simeq y_e^{(2)}
\begin{pmatrix}
0 & - \sin \theta _{eL} \sin \theta _{eR} \sin \left(\zeta _{e_L}-\omega _{eL}\right) & - \cos \theta _{eR} \sin \theta _{eL} \sin \left(\zeta _{e_L}-\omega _{eL}\right) \\
0 &   \sin \theta _{eL} \sin \theta _{eR} \cos \left(\zeta _{e_L}-\omega _{eL}\right) &  \cos \theta _{eR}  \sin \theta _{eL} \cos \left(\zeta _{e_L}-\omega _{eL}\right) \\
0 & \cos \theta _{eL}  \sin \theta _{eR} &  \cos \theta _{eL} \cos \theta _{eR} \\
\end{pmatrix}\;.
\end{equation}

For generic values of the high energy mixing angles, the elements of the second and third column of this matrix will have comparable entries. We therefore consider for concreteness the  process $\mu\rightarrow e\gamma$, which is the most strongly constrained by experiments. For a wide range of parameters, the largest contribution to this process are  two loop Barr-Zee diagrams~\cite{Paradisi:2005tk,Hisano:2010es} with a top-quark in the loop. Assuming $m_H\gg v$, the branching ratio is given by:
\begin{equation}
{ \rm BR} (\mu \rightarrow e\, \gamma) \simeq
  \frac{8 \alpha^3}{3 \pi^3}
  \frac{|{\tilde Y}_{e12}^{(2)}|^2}{|{\tilde Y}_{e22}^{(1)}|^2}
  \left| \frac{|\lambda_6|v^2}{m_H^2}
f \!\! \left(\frac{m_t^2}{m_h^2} \right) -
  \frac{Y_{u33}^{(2)}}{ Y_{u33}^{(1)}} \frac{m_t^2}{m_H^2} \log^2 \frac{m_t^2}{m_H^2} \right|^2\;.
  \label{eq:BRmuegamma}
\end{equation}
Here, $m_t$ denotes the top quark mass, $Y_{u33}^{(a)}$ is the Yukawa coupling of the top quark to the Higgs $\Phi_a$ (expressed in the basis where $Y_u^{(1)}$ is diagonal), $\lambda_6$ is the coupling constant of the potential term $V\supset  \lambda_6 (\Phi_1^\dagger \Phi_1) (\Phi_1^\dagger \Phi_2) $ and $f(z)$ is a function defined in \cite{Hisano:2010es} which evaluates $f(2)\approx 1$. Assuming again $y_{e,u}^{(2)}\sim y_{e,u}^{(1)}$ and $\lambda_6\sim 1 $, one finds
\begin{equation}
{ \rm BR} (\mu \rightarrow e\, \gamma) \sim 10^{-14} \left(\frac{m_H}{10\,{\rm TeV}}\right)^{-4}\;.
\end{equation}
Therefore, the non-observation of the process $\mu\rightarrow e\,\gamma$ typically requires the particles of the extended Higgs sector to have masses larger than  $\sim $10 TeV, unless the couplings $y_{e,u}^{(2)}$ and $\lambda_6$ are suppressed. A large mass scale for the exotic Higgs states, however,   barely affects the leptonic mass ratios and mixing angles calculated in section \ref{section:quantum-effects}, which depend   at most logarithmically on $m_H$.

\section{Conclusions}
\label{section:conclusions}

We have studied the impact of quantum effects on the leptonic parameters in the Standard Model extended by one right-handed neutrino and one extra Higgs doublet. No additional discrete or global symmetry was assumed. We have shown that, starting with rank-1 Yukawa matrices at a cut-off scale,  quantum  effects generate  a mass for the muon and for the next-to-heaviest neutrino which are in qualitative agreement with the experimental data. The radiatively induced mass hierarchies mostly depend on the eigenvalues and flavour structure of the tree level Yukawa matrices and, in the case of the neutrino mass hierarchy, also on the value of the quartic coupling $\lambda_5$. The dependence on the mass scales of the model is, on the other hand, only logarithmic and hence fairly mild. Furthermore, in this framework it is generically expected a leptonic mixing matrix with all entries ${\cal O}(0.1)$, namely with mixing angles that are neither maximal nor small, also in qualitative agreement with neutrino oscillation experiments. Similar conclusions are expected in models with more right-handed neutrinos and/or rank-3 Yukawa matrices, provided the tree level mass hierarchies are much larger than the radiatively generated ones.  This framework contains new sources of lepton flavour violation that induce rare leptonic decays, such as $\mu\rightarrow e\gamma$, with a rate that depends on the inverse mass of the extra Higgs states to the fourth power. On the other hand, the fermionic mass ratios depend only logarithmically on this mass, therefore, by sufficiently increasing the mass of the exotic scalar states, it is possible to suppress the rates for the rare decays without significantly affecting the mechanism of generation of the fermionic mass hierarchies and mixing angles discussed in this paper.

\section*{Acknowledgements}

This work was supported in part by the DFG cluster of excellence ``Origin and Structure of the Universe'' and by the ERC Advanced Grant project ``FLAVOUR''(267104) (A.I.).

\appendix
\section{Beta functions}
\label{app:Beta-functions}
The one-loop $\beta$ functions of the Yukawa couplings $Y^{(a)}_{e,\nu}$ are defined from the renormalization group equations
\begin{equation}
\frac{d Y^{(a)}_{e,\nu}(\mu)}{d\mu}=\frac{1}{16\pi^2}\beta_{e,\nu}^{(a)}(\mu)\;,
\end{equation}
where $\mu$ is the renormalization scale. They were calculated for the multi-Higgs doublet model in \cite{Grimus:2004yh} and read:
\begin{eqnarray}
  \beta_e^{(a)}(\mu)&=& 
	\left(-\frac{9}{4}g^2-\frac{15}{4}g'^2 \right) Y_e^{(a)}
	+\left[	
		3{\rm Tr}\left(Y_u^{(a)\dagger}Y_u^{(c)}+Y_d^{(a)}Y_d^{(c)\dagger}\right)
		+{\rm Tr}\left(Y_e^{(a)}Y_e^{(c)\dagger}\right)
	\right]Y_e^{(c)}  \nonumber  \\ 
&& +Y_e^{(a)}Y_e^{(c)\dagger}Y_e^{(c)} +\frac{1}{2}Y_e^{(c)}Y_e^{(c)\dagger}Y_e^{(a)}  \nonumber \\ 
&&+\left\{{\rm Tr}(Y_\nu^{(a)\dagger}Y_\nu^{(c)}) Y_e^{(c)} 	-2Y_\nu^{(c)}Y_\nu^{(a)\dagger}Y_e^{(c)} +\frac{1}{2}Y_\nu^{(c)}Y_\nu^{(c)\dagger}Y_e^{(a)}\right\}\Theta(\mu-M_{\rm maj}) ,\\
        \beta_\nu^{(a)}(\mu)&=&
	\left\{\left[-\frac{9}{4}g^2-\frac{3}{4}g'^2 \right] Y_\nu^{(a)}
	+\left[	
		3{\rm Tr}(Y_u^{(a)}Y_u^{(c)\dagger}\!+\!Y_d^{(a)\dagger}Y_d^{(c)})
		+{\rm Tr}(Y_\nu^{(a)}Y_\nu^{(c)\dagger}+Y_e^{(a)\dagger}Y_e^{(c)})
	\right]Y_\nu^{(c)} \right.\nonumber\\
	&& \left.
	-2Y_e^{(c)}Y_e^{(a)\dagger}Y_\nu^{(c)}
	+Y_\nu^{(a)}Y_\nu^{(c)\dagger}Y_\nu^{(c)}
	+\frac{1}{2}Y_e^{(c)}Y_e^{(c)\dagger}Y_\nu^{(a)}
	+\frac{1}{2}Y_\nu^{(c)}Y_\nu^{(c)\dagger}Y_\nu^{(a)}\right\}\Theta(\mu-M_{\rm maj}) \nonumber\\
 \end{eqnarray}
where summation over repeated indices is understood and $\Theta(x)$ is the Heaviside function, which takes into account the fact that for $\mu < M_{\rm maj}$ the right-handed neutrinos are decoupled. 

For energy scales below the right-handed neutrino Majorana mass, the theory is described by the charged lepton Yukawa coupling $Y_e^{(a)}$ and by the dimension five operators $\kappa^{(ab)}$. The renormalization group equations for the couplings $\kappa^{(ab)}$ are
\begin{equation}
\frac{d\kappa^{(ab)}}{d\mu}=\frac{1}{16\pi^2}\beta_\kappa^{(ab)}(\mu)
\end{equation}
where the corresponding beta functions were also calculated in \cite{Grimus:2004yh} and read:
\begin{eqnarray}
 \beta_\kappa^{(ab)}(\mu)&=&
	\frac{1}{2} \left[ 
		Y_e^{(c)}Y_e^{(c)\dagger} \kappa^{(ab)}+
		\kappa^{(ab)} \left(Y_e^{(c)}Y_e^{(c)\dagger}\right)^T 
	\right]
	+2 \left[
		Y_e^{(c)}Y_e^{(b)\dagger} \kappa^{(ac)} +
		\kappa^{(cb)} \left( Y_e^{(c)}Y_e^{(a)\dagger} \right)^T 
	\right]\nonumber\\
	&&
	-2 \left[ 
		Y_e^{(c)}Y_e^{(a)\dagger} (\kappa^{(cb)} + \kappa^{(bc)}) +
		(\kappa^{(ac)}+\kappa^{(ca)}) \left( Y_e^{(c)}Y_e^{(b)\dagger} \right)^T 
	\right]\nonumber\\
	&&
	 + \left[ 3 {\rm Tr}(Y_u^{(a)}Y_u^{(c)\dagger} + Y_d^{(a)\dagger}Y_d^{(c)}) +
		{\rm Tr}(Y_e^{(a)\dagger}Y_e^{(c)})
			 \right] \kappa^{(cb)}\nonumber\\
	&&
	 +\kappa^{(ac)} \left[ 3 {\rm Tr}(Y_u^{(b)}Y_u^{(c)\dagger} + Y_d^{(b)\dagger}Y_d^{(c)})+
		{\rm Tr}(Y_e^{(b)\dagger}Y_e^{(c)})
			 \right]\nonumber\\
	&&
	 -3 g^2 \kappa^{(ab)}
	 +2 \lambda_{acbd} \kappa^{(cd)}\;,
\end{eqnarray}
where the quartic couplings $\lambda$ are defined by
$V\supset \frac{1}{2}\lambda_{abcd}(\Phi^\dagger_a \Phi_b)(\Phi^\dagger_c \Phi_d)$.

\section{Functions $P$ and $Q$}
\label{app:P&Q}

The functions $P$ and $Q$, defined in  Eqs.~(\ref{eq:def_p&q}), determine the ratio between the muon and the tau mass, Eq.~(\ref{eq:ratio-approx}), as well as the contribution to the leptonic mixing from the charged lepton sector, Eq.~(\ref{eq:zetaeL}). The functions $P$ and $Q$ are linear combinations of the functions $p_x$ and $q_x$, $x=e,\nu,u,d$, which are explicitly given by:
\begin{eqnarray}
p_{e} &=& \frac{1}{4} \sin2 \theta _{eL} \sin 2 \theta _{eR}\sin \omega _{eL} \;, \nonumber \\
q_{e} &=& \frac{1}{4} \sin2 \theta _{eL} \sin 2 \theta _{eR}\cos \omega _{eL} \;,
\end{eqnarray}
\begin{eqnarray}
p_{\nu} &=& -2\cos \theta _{eL}\sin\theta_{e_R}\cos \alpha  \sin \theta _{\nu } \sin \omega _{\nu }\nonumber\\
&&+\sin \theta _{eL} \sin\theta_{e_R}\left(\cos \alpha  \sin \omega _{eL} \cos \theta _{\nu }+\sin \alpha  \sin \theta _{\nu } \left(\sin \omega _{eL} \cos \omega _{\nu }-2 \cos \omega _{eL} \sin \omega _{\nu }\right)\right)  \;,\nonumber\\
q_{\nu} &=& -2\cos \theta _{eL} \sin\theta_{e_R}  \cos \alpha  \sin \theta _{\nu } \cos \omega _{\nu } \nonumber\\
&&+\sin \theta _{eL} \sin\theta_{e_R}\cos \omega _{eL} \left(\cos \alpha  \cos \theta _{\nu }-\sin \alpha  \sin \theta _{\nu } \cos \omega _{\nu }\right) \;,
\end{eqnarray}
\begin{eqnarray}
p_{ u}&=&3 \cos \theta_{uL} \cos \theta_{uR} \sin \theta_{e_L}\sin \theta_{e_R} \sin  \omega _{eL}  \;,\nonumber\\
q_{ u}&=&3 \cos \theta_{uL} \cos \theta_{uR} \sin \theta_{e_L}\sin \theta_{e_R} \cos \omega_{eL} \;,
\end{eqnarray}
\begin{eqnarray}
p_{ d}&=&3 \cos \theta_{dL} \cos \theta_{dR} \sin \theta_{e_L}\sin \theta_{e_R} \sin  \omega _{eL} \;, \nonumber\\
q_{ d}&=&3 \cos \theta_{dL} \cos \theta_{dR} \sin \theta_{e_L}\sin \theta_{e_R} \cos \omega_{eL} \;.
\end{eqnarray}

\end{document}